\documentclass[twocolumn]{aa}
\usepackage{graphicx}
\usepackage{txfonts}
\begin{document}
   \title{Extinction properties of the X-ray bright/optically faint
afterglow of GRB 020405}


   \author{G. Stratta
          \inst{1},
          R. Perna\inst{2,4}, D. Lazzati\inst{2}, F. Fiore\inst{3}, 
	L.A. Antonelli\inst{3}, M.L. Conciatore\inst{3}
         }

   \offprints{G. Stratta}

   \institute{LATT, Laboratoire d'Astrophysique Toulouse-Tarbes,
              14, av. Edouard Belin, 31400 Toulouse, France\\
         \and
             JILA and Department of Astrophysical and Planetary Sciences,
	     University of Colorado at Boulder, 440 UCB, 
	     Boulder, CO, 80309, USA\\
	\and
	     INAF, Osservatorio Astronomico di Roma, 
	     via Frascati 33, Monteporzio Catone, 00040, Rome, Italy\\ 
	\and
	    Department of Astrophysical Sciences, 4 Ivy Lane, Princeton, NJ, 08544, USA\\
		}

   \date{Received ; accepted }

   \abstract{
We present an optical-to-X-ray spectral analysis of the afterglow of
GRB~020405. The optical spectral energy distribution not corrected for
the extragalactic extinction is significantly below the X-ray
extrapolation of the single powerlaw spectral model
suggested by multiwavelength studies.  We investigate whether considerable extinction could
explain the observed spectral ``mismatch'' by testing several types of
extinction curves.  For the first time we test extinction curves
computed with time-dependent numerical simulations of dust grains
destruction by the burst radiation.  We find that an extinction law weakly dependent on
wavelength can reconcile the unabsorbed optical and X-ray data with
the expected synchrotron spectrum.  A gray extinction law can be
provided by a dust grain size distribution biased toward large grains.
   }

   \authorrunning{G. Stratta et al.}

   \maketitle
%

\section{Introduction}

Long-duration Gamma Ray Bursts (GRBs) are associated with the core
collapse of massive stars exploding as type Ic supernovae (GRB 980425,
Galama et al.  1998; GRB 011211, Della Valle et al. 2003; GRB 030329,
Stanek et al.  2003; GRB 031203, Malesani et al.  2004).  Due to their
short life-time, massive stars are likely to die within their
birthplace.  Consequently, long-duration GRBs occur in the same star
formation region where their progenitor star was born and where it
rapidly evolved. If high redshift galaxy star formation regions are
akin to Galactic giant molecular clouds, a dense and dusty environment
is expected in the vicinity of GRBs.
 
Indeed, the measured equivalent hydrogen column densities $N_H$
obtained from X-ray afterglow spectral analysis are on average
consistent with those observed along the line of sight of Galactic
molecular clouds (e.g. Galama \& Wijers 2001, De Pasquale et al. 2003,
Stratta et al. 2004).  In some afterglows (e.g. GRB 971214,
Ramaprakash et al. 1998; GRB 980703 Bloom et al.  1998; GRB 980329,
Yost et al.  2002) a high visual extinction $A_V$ in the GRB host
galaxy has been measured.  In addition, a large dust content was
inferred from high resolution spectroscopy by refractory metal
abundance analysis along the line of sight of four optical afterglows
(Savaglio et al. 2003, 2004).  However, on average, the estimated
afterglow reddening is low and the rest frame visual extinction is
a factor 10-100 lower than expected from the extrapolation of
the X-ray $N_H$, assuming the Galactic dust-to-gas ratio (\v{S}imon et
al. 2001; Galama \& Wijers 2000). In a few high redshift GRBs for which
high resolution afterglow spectroscopy was performed, high $N_H$ from
$Ly_{\alpha}$ absorption was measured. On the other hand a low $A_V$
is inferred from continuum absorption modeled with the Small
Magellanic Cloud (SMC) extinction law.  These results can be explained
by a low metallicity and/or a low dust-to-gas ratio in the burst
environment (e.g. Hjorth et al. 2003, Vreeswijk et al. 2004).
 
Alternatively, the GRB environment may be characterized by a
``non-standard'' extinction law (neither Galactic nor SMC; Stratta et
al. 2004, Savaglio et al. 2004).  In particular, an extinction law
weakly dependent on wavelength can provide high visual extinction
without substantial reddening (e.g. Hjorth et al. 2003).  Dust
properties of high redshift environments - such as the GRB host galaxies - 
are still poorly known.  We have some
indications from the Magellanic Clouds (e.g. Pei et al. 1992) and from
a sample of local starburst galaxies (e.g. Calzetti et al. 1997) and
Active Galactic Nuclei (e.g. Maiolino, Marconi \& Oliva 2001, Maiolino 2004) that the dust
composition, grain size distribution and dust to gas ratio can be
significantly different from the ones observed in our Galaxy.  The GRB
star-forming environment, in addition, may not be representative of
the typical low-density host galaxy inter-stellar matter (ISM). Finally, the intense X-ray
and UV flux from the burst itself can modify the intrinsic dust grain
size distribution by differential dust grain destruction, since
sublimation processes are more effective on small grains (e.g. Waxman
\& Draine 2000; Fruchter et al. 2001; Draine \& Hao 2002; Perna \&
Lazzati 2002; Perna, Lazzati \& Fiore 2003).
                                                                                
A diagnostic tool of the GRB environment extinction properties is the simultaneous
optical-to-X-ray afterglow continuum spectral analysis.  The intrinsic
optical continuum can be extrapolated from the X-ray data and dust
absorption can be measured from the deviations of the measurements
from the expected fluxes. Good quality data are necessary to constrain
the free parameters of the continuum model (the position of the
cooling break possibly laying between the optical and X-ray
bands). This method was applied, with a various degree of accuracy, in
a handful of GRBs (Stratta et al. 2004).
 
In this work we present a study of the optically faint/X-ray bright
afterglow of GRB 020405 (e.g. Dado et al. 2002, Masetti et al. 2003,
Bersier et al. 2003, Covino et al. 2003, Price et al. 2003, Berger et
al. 2003, Mirabal et al. 2003).  This burst was discovered by the IPN
on 2002 April 5.029 UT. The duration of the burst was $\sim40$s, the
25-100 keV fluence was $\sim3\times10^{-5}$ergs cm$^{-2}$ and the peak
flux was $\sim10^{-6}$ ergs cm$^{-2}$ s$^{-1}$. This GRB was also
observed by the GRB Monitor on board BeppoSAX with a duration of
$\sim60$s in the 40-700 keV band and a 50-700 keV fluence of
$4\times10^{-5}$ ergs cm$^{-2}$ s$^{-1}$ (e.g. Price et al. 2003).
Optical observations started 18 hours after the burst event (Price et
al. 2003) for a period of $\sim10$ days. An unknown fading source was
localized at R.A.13$^{h}$ 58$^{m}$ 03$^{s}$.12 and Dec. -31$^{\circ}$
22$'$ 22$''$.2 with an uncertainty of 0$''$.3 (Masetti et al. 2002, Price et
al. 2003).  An exhaustive summary of all the afterglow observations
performed in the optical band is presented in Masetti et al. (2003).
The host galaxy has redshift $z=0.691\pm0.002$ (Masetti et al. 2003).  
Chandra observations started on April 6.711 UT and lasted
until 7.350 UT using LETGS in conjunction with the ACIS detector and
revealed a new fading source that was identified as the X-ray
afterglow counterpart.  Mirabal et al. (2003) found a featureless
spectrum, well described by a power law continuum with energy spectral
index $\alpha_X=0.72\pm0.21$ and a rest frame $N_{H}$ of
$(4.7\pm3.7)\times10^{21}$cm$^{-2}$. The X-ray light curve decayed as
a power law, with a temporal decay index $\delta_X=1.97\pm1.10$
(Mirabal et al. 2003). The optical (VRI) decay index is 
$\delta_O=1.54\pm0.06$ from 1 to 10 days after the burst (Masetti et al. 2003).  
The NIR bands (J and H) show a less steep decay with
index $\delta_{IR}\sim1.3$ (Masetti et al. 2003). 
Radio observations show a rapidly fading
"radio flare" at 1.2 days from the burst event (Berger et al. 2003).
Berger et al. (2003) found that the radio, optical and X-ray data are 
self-consistently modeled assuming a collimated ejecta expanding into a uniform
medium. Best fit parameters yield a jet break time at about 1 day after
the burst and indicate that the cooling frequency at that time is at energies 
lower than the optical band. 
The same conclusion on the location of the 
cooling frequency was obtained by Masetti et al. (2003), 
who showed how the optical spectral energy distribution has a break around the J band.
For these reasons, for this burst there is convincing evidence
for the absence of a spectral break between the X-ray and the optical
band.  The different decay index and spectral slope observed in the
NIR band with respect to the optical may be due to the presence of the cooling
frequency between the optical and the NIR bands (Masetti et al. 2003; Berger et 
al. 2003).
 
The paper is organized as follows: in \S 2 we summarize the data
analysis procedure; in \S 3 we describe the adopted extinction curves;
in \S 4 and \S 5 we present and discuss our results.

%

\begin{table*}
\caption{The optical-near infrared magnitudes of the optical afterglow of GRB 020405 
(from Masetti et al. 2003), corrected for Galactic extinction and rescaled
  at $t_0$=UT 7.01 ($T_{GRB}$+1.98 days)}            
\label{tab:1}      
\centering                          
\begin{tabular}{cccccccc}        
\hline\hline                 

U & B          & V      & R      & I  & J & H & K   \\
\hline
 $21.22\pm0.12$ & $21.90\pm0.1$ & $21.50\pm0.08$ & $21.08\pm0.06
$ & $20.56\pm0.12$ & $19.37\pm0.10$ & $18.72\pm0.10$& $17.99\pm0.10$ \\
\hline                                   
\end{tabular}
\end{table*}


\section{Data analysis procedure}

We followed the standard procedure for the X-rays (0.1-10.0 keV) data
reduction for Chandra ACIS data.  We grouped energy channels in order
to have at least 20 counts per bin in order to apply the $\chi^2$
statistic in the fitting procedure.  We first fit the X-rays spectrum
with a power law model and two absorber components. 
The column density of the first absorber
was fixed to the Galactic value toward the line of sight of the burst
$4.3\times10^{20}$cm$^{-2}$ (Dickey \& Lockmann 1995). We subsequently
allowed for the second absorber a free column density to constrain 
a possible contribution of absorption from the host galaxy ISM.  
We found that the data are well fitted by this model, with a best fit 
energy spectral index of $\alpha_X=1.0\pm0.2$ and rest frame ($z=0.691$) 
hydrogen column density of $N_{Hz}=(0.8\pm0.2)\times10^{22}$cm$^{-2}$ 
in addition to the Galactic absorption ($\chi^2$/d.o.f.=22.3/24, 
where d.o.f. stands for degrees of freedom). The best fit model (absorbed power law)
1.6-10.0 keV flux is $8.7\times10^{-13}$ ergs cm$^{-2}$ s$^{-1}$.
Errors are at 1 $\sigma$ level.  We note that our spectral index is
steeper than, but still consistent with, the one obtained by Mirabal
et al. (2003).
 
The optical-NIR SED was obtained by taking the photometric points from
the journal table published by Masetti et al. (2003).  Observations in
each photometric band (U,B,V,R,I,J,H and K) were performed at
different times and with different telescopes.  We extrapolated the
magnitudes at the observational epoch $t_0$ of 7.01 UT (1.98 days
after the burst). To this purpose, we firstly selected the
observations performed as closer as possible to $t_0$, and then we
extrapolated the magnitude at $t_0$ assuming a single powerlaw with
decay index estimated at that time from Masetti et al. (2003). The
decay index uncertainty was propagated to the magnitude errors with an
average increase that depends on the temporal distance of the
photometric measure from $t_0$ (see Tab.1 in Masetti et al. 2003).
This results in an increase of the error on the magnitudes of 20\% or more, 
depending on the temporal distance from the selected $t_0$.
We choose $t_0$ as the average time of the X-ray observations given
the large uncertainty in the X-ray decay slope (Mirabal et al. 2003).
 
We corrected the magnitudes for Galactic extinction (E(B-V)=0.054 mag)
using the Galactic dust infrared map by Schlegel et al. (1998).
Assuming a total-to-selective extinction $R_V=A_V/E(B-V)=3.1$, we
derived the extinction value in each photometric band from the
Galactic extinction curve parameterization $A(\lambda)/A_V$ by
Cardelli et al. (1989).  We finally converted magnitudes to fluxes
using the effective wavelengths and normalization fluxes given by
Fukugita et al. (1995).  
A $7\%$ systematic error was added quadratically to the magnitude
errors to account for mag-to-flux conversion uncertainties and for 
intercalibration errors from different telescopes (see also Masetti et al. 2001). 
The resulting magnitudes are summarized in Table~\ref{tab:1}.
  
A {\it simultaneous} fit of the NIR-to-X-rays spectrum was performed
with the X-ray spectrum corrected for the total photoelectric
absorption and the optical SED corrected for the Galactic extinction
but not corrected for an extra-Galactic component. 
For the intrinsic SED, we considered both a single powerlaw
and a broken powerlaw, the latter with break energy below the optical range.  
In order to estimate the host galaxy extinction we multiplied the continuum
spectral model by an absorption component.  We tested different ISM
dust composition and dust to gas ratios by assuming several type of
extinction curves (see \S 3).


   \begin{figure}
   \centering
   \includegraphics[width=9cm]{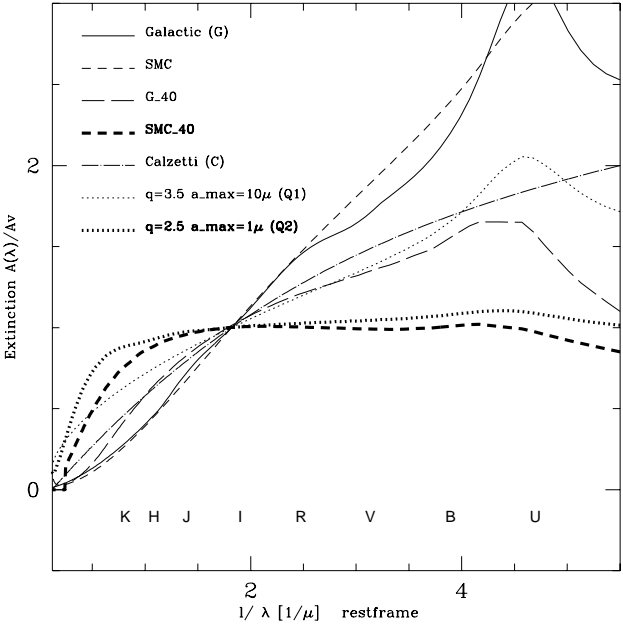}
   \caption{Extinction curves adopted in the analysis of the optical-NIR
 photometry against $1/\lambda$ where $\lambda$ is the rest frame wavelength.  
 We note that the weakest wavelength
 dependence is shown by the Q2 extinction curve and by the
 SMC-modified extinction curve computed by dust destruction numerical
 simulations (SMC\_40, Perna et al. 2003).  We also indicate the
 position of the photometric filters for a system located at
 $z=0.695$, the redshift of GRB~020405.  Extinction curves that
 provide an acceptable fit to the data are emphasized with thicker
 lines.}
              \label{fig:curves}%
    \end{figure}

\section{Host galaxy extinction}
 
Since we do not know a priori the host galaxy extinction curve, we
tested several types of environments (Fig.~\ref{fig:curves}) with
different dust grain size distributions, dust compositions and dust to
gas mass ratios.  The ratio between the hydrogen column density $N_H$
and the visual extinction $A_V$ provides a measure of the dust-to-gas
ratio for a given dust distribution and composition. 
For simulated time-dependent extinction laws, the dust-to-gas ratio at 
the relevant time is derived from the simulations.
 
We tested the following extinction curves:
\begin{itemize}
\item the Galactic extinction curve (hereafter G) from Cardelli et
al. (1989), for which $N_H/A_V=0.18\times10^{22}$cm$^{-2}$ (Predehl \&
Schmidt 1995).
\item
The Small Magellanic Cloud extinction curve (hereafter SMC) from Pei
et al. (1992), for which $N_H/A_V=1.6\times10^{22}$cm$^{-2}$
(Weingartner \& Draine 2000).
\item
The attenuation curve derived for a sample of local starburst galaxies
(hereafter C) from Calzetti et al. (1994). No $N_H/A_V$ has been
measured for this case due to the complexity of the geometry of the
dust and star distribution inside these galaxies (Calzetti et
al. 2001).
\item
Two extinction curves obtained by Maiolino et al. (2001) for the
environment of a sample of AGNs. The Q1 extinction curve is computed
assuming a power law dust grain size distribution $dn(a)\propto a^{-q}da$ in the
range $a_{min}=0.005\mu < a <a_{max}=10\mu$ and $q=-3.5$. In this case
$N_H/A_V=0.7\times10^{22}$ cm$^{-2}$. The Q2 extinction curve is
derived assuming $a_{min}=0.005\mu < a <a_{max}=1\mu$ and $q=-2.5$ and
it yields $N_H/A_V=0.3\times10^{22}$ cm$^{-2}$.
\item
Extinction curves resulting from numerical simulations performed with
the code of Perna \& Lazzati (2002). This code computes the temporal
evolution of the dust grain size distribution, taking into account
grain erosion due to UV and X-ray illumination. The ionizing continuum
is tailored to the case of GRB~020405 (see above). 
Only the 40 seconds of the burst emission were considered,
since the overall fluence of the later afterglow was smaller 
than that of the prompt emission by at least an order of
magnitude. Initial conditions
were either a Galactic (final curve labelled G\_40) or SMC (final
curve labelled SMC\_40) ISM. A uniform cloud with radius
$R=7\times10^{22}$~cm and density $n=4\times5$~cm$^{-3}$ was
considered. We checked that changing this set-up would not change our
conclusions. 
\end{itemize}


   \begin{figure}
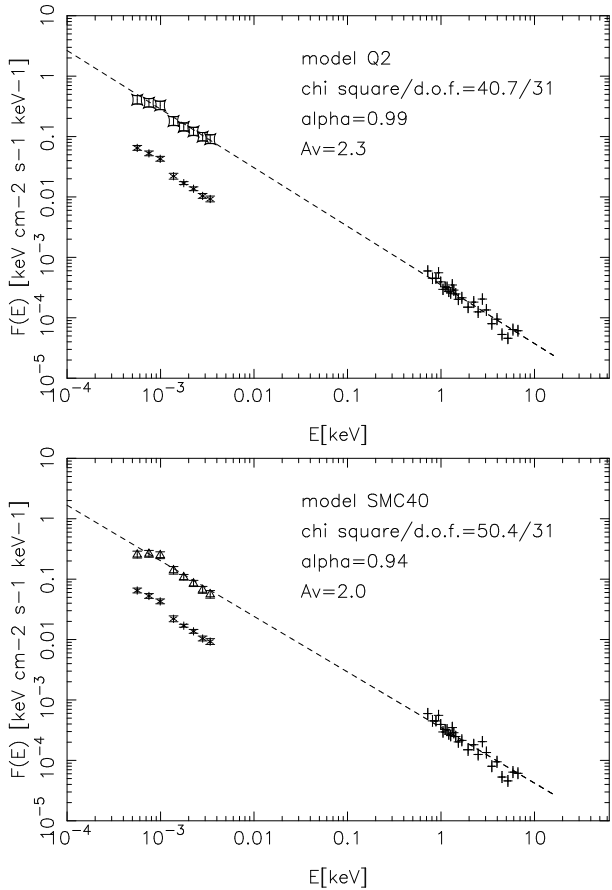

   \centering
   \includegraphics[angle=-90,width=8cm]{3302fig2.ps}
   \includegraphics[angle=-90,width=8cm]{3302fig3.ps}
   \caption{NIR-to-X-ray absorbed power-law fits to the data of
GRB~020405.  A Q2 absorption curve is used in the upper panel, while a
SMC\_40 curve is used in the bottom panel. Optical data are shown both
uncorrected (asterisks) and corrected (squares and triangles) for the
host galaxy extinction.
See Table~\ref{tab:2} for the best fit parameters.}
              \label{fig:fit1}%
    \end{figure}

\section{Results}
 
We fit the NIR/optical and the X-ray spectral energy distribution,
extrapolated to a common epoch (see \S 2), with a dust-absorbed power
law model.  The NIR-to-X-ray spectral index $\alpha$ was let free to vary.
We found that the optical/NIR data corrected for the
``standard'' extinction curves, namely the G and the SMC, are
incompatible with X-ray data (see Tab.~\ref{tab:2}). Unacceptable fits were
obtained also with the Calzetti, the Q1 and the G\_40 curves (see
\S3).
 
Acceptable fits with comparable probability $P(>\chi^2)$ of $15\%$,
were instead obtained with the Q2 and the SMC\_40 curves.  These
curves have the weakest dependence on wavelengths among those
considered (see Fig.~\ref{fig:curves}).  We note that the relatively
high $\chi^2$ levels are due to the apparent bend of the continuum in
the NIR which cannot be reproduced by a power-law
(Fig.~\ref{fig:fit1}).
 
We tested therefore a smoothed broken power law spectrum (with the
sharpness parameter $s$ equal to 1, Granot \& Sari 2002) with break
energy between the optical and NIR bands.  The optical-to-X-rays
spectral index was let free to vary within the 90$\%$ confidence level
interval 0.8-1.0, found from the X-ray analysis.  Assuming
$\nu_c<\nu_X,\nu_O$, from the X-ray spectral slope we derive an
electron spectral index $p=2.0\pm0.4$. We then compute the expected
NIR spectral slope (corresponding to $(p-1)/2$ for $\nu_{NIR}<\nu_c$,
Sari et al. 1998) that we let free to vary within the 90$\%$
confidence level interval ($0.3-0.7$).  Even allowing for this extra
degree of freedom, we found that the G, SMC, C, Q1 and G\_40
absorption models are inconsistent with the data (see
Tab.~\ref{tab:3}).
 
\begin{figure}
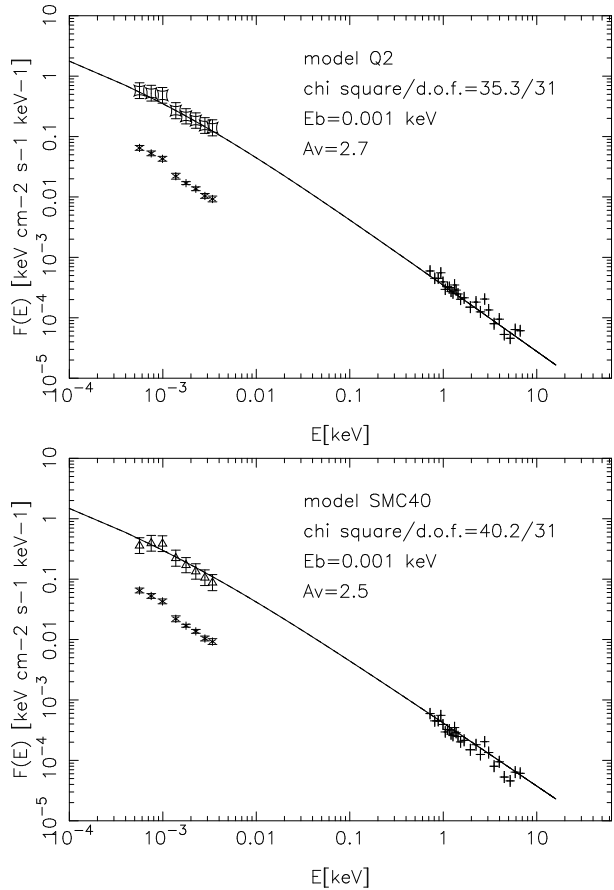

   \centering
   \includegraphics[angle=-90,width=8cm]{3302fig4.ps}
   \includegraphics[angle=-90,width=8cm]{3302fig5.ps}
   \caption{Same as Fig.~\ref{fig:fit1} but with a smoothed broken
power-law continuum model (see Table~\ref{tab:3}). }
              \label{fig:fit2}%
    \end{figure}

For the models Q2 and SMC\_40 we find an improvement in the fit with a
$2.5\sigma$ statistical significance, according to the F-test
(Fig.~\ref{fig:fit2}).  They both fit the data successfully
(Tab.~\ref{tab:3}).  We then checked the consistency of the measured
$A_V$ values with the $N_{Hz}$ measured from the X-ray analysis and
the $N_H/A_V$ relationship expected (see \S 3).  For the Q2 model, we
found $N_H=0.7\times10^{22}$cm$^{-2}$ assuming a power law model and
$N_H=0.8\times10^{22}$cm$^{-2}$ assuming a broken power law.  These
$N_H$ values are consistent with the $N_{Hz}$ [($0.8\pm0.2)\times
10^{22}$ cm$^{-2}$] measured from X-ray analysis.  For the SMC\_40
model, even though the shape of the extinction curve fits the data
well, a gray extinction with $A_V\gtrsim2$ is obtained only for
$N_{H}\sim10^{24}$ cm$^{-2}$, largely exceeding the X-ray measured
$N_{Hz}$. Formally a self consistent fit can be obtained assuming a
dust-to-gas ratio 100 times larger than the SMC one. This would imply
a $\gtrsim10$ times solar metallicity. We consider such physical
conditions too extreme, even tough GRB (star formation) environments
are expected to be especially dusty.
 
\section{Discussion}
 
We have studied the broad-band spectrum of the afterglow of GRB~020405
from NIR to X-ray bands. The optical-NIR and X-ray spectral indices
are mutually consistent. However, this burst is peculiar since it has
an optical-NIR flux a factor of $\gtrsim10$ times dimmer than the
extrapolation of the X-ray spectrum. This can be attributed to an
inverse Compton component in the X-ray band (Sari \& Esin 2001).
Alternatively, the optical brightness and spectrum could be affected
by dust extinction. We investigated the latter possibility.  We found
a self-consistent model that allowed us to constrain the properties 
of the dust distribution and its formation history.
 
From a simultaneous NIR-to-X-ray spectral analysis, we found that the
optical data can be fit with the external shock model only if
corrected for an extinction curve weakly dependent on wavelength. Such
an extinction curve can be the result of small dust grains coagulation
and/or dust grain destruction (e.g. Maiolino et al. 2001, Perna et
al. 2003).  In Figure~\ref{fig:curves} we plot the extinction curves
derived from these two different processes (see \S3).  The remarkable
similarity of these extinction curves, computed assuming different
processes, is evident.
 
The dust coagulation rate increases with density $\propto n^{1/2}$ (Draine 1985). 
It is therefore favored in dense environments such as the cores of star
formation regions where long GRBs are expected to
explode. Incidentally, extinction curves weakly dependent on
wavelengths have been inferred also for other extra-galactic objects,
such as AGNs (Maiolino et al. 2001).
 
A dust distribution biased towards large grains can also be due to
grain destruction mechanisms (Waxman \& Draine 2000; Fruchter et
al. 2001; Draine \& Hao 2002; Perna \& Lazzati 2002). We have
attempted to reproduce the observations starting from known dust
distributions in the local Universe (Galactic and SMC). The evolution of the
initial distributions is followed as a result of the grain interaction
with the burst ionizing flux (Perna et al. 2003). We find that the
modified Galactic distribution cannot reproduce the observations, while the
SMC one can reproduces them if an extremely large dust-to-gas ratio
and metallicity are assumed. Since these conditions are hardly
realized in nature, a scenario in which the large grain distribution
is intrinsic to the host galaxy is favored.
 
The difficulty in reproducing the results as a consequence of dust
destruction can be understood since dust destruction reduces the
opacity besides modifying the reddening law. Since this burst requires
a large extinction on a day timescale, it is likely that the opacity
is provided by dust at a large distance from the burster, which is not
affected by the burst photons. Prompt optical-NIR observations are
better suited for the detection of dust destruction, since the
evolution is caught ``on the act''. This would allow us to detect dust
before it is destroyed and, based on the properties of what is left,
infer the geometrical and physical conditions of the host galaxy ISM
(Perna et al. 2003).

\begin{table}
\begin{tabular}{lccc}
\hline
\hline
Extinction & $\alpha$ & $A_V$ & $\chi^2/d.o.f.$  \\
curve      &       & mag   &       \\
\hline
  G   & $0.75\pm0.01$  & $0.36\pm0.04$ & 89.3/31 \\
\hline
  SMC  & $0.74\pm0.01$  & $0.33\pm0.04$ & 83.8/31 \\
  SMC\_40   & $0.94\pm0.02$ & $2.0\pm0.1$  & 50.4/31 \\
\hline
  C    & $0.80\pm0.02$  &  $0.78\pm0.08$& 66.5/31 \\
  Q1   & $0.79\pm0.01$  & $0.69\pm0.08$ & 76.7/31 \\
  Q2   & $0.99\pm0.02$ & $2.27\pm0.14$  & 40.7/31 \\
\hline
\end{tabular}
\caption{Best fit parameters from simultaneous NIR-to-X-ray spectral
analysis assuming a power law model and different
extinction curves to evaluate rest frame host galaxy extinction $A_V$
(see \S3)
\label{tab:2}}
\end{table}

\begin{table}
\begin{tabular}{lccccc}
\hline
\hline
Extinction & $E_{break}$ & $A_V$ & $\chi^2/d.o.f.$  \\
 curve     & $\mu$ & mag   &       & \\
\hline
G   & not found & -  & -  \\
\hline
 SMC    & not found & -  & - \\
 SMC\_40  &  $1.12\pm0.20$ & $2.53\pm0.3$ & 40.2/31 \\
\hline
  C     & $1.3\pm0.5$  & $0.87\pm0.11$ & 64.5/31      \\
  Q1    & not found    & -              & -       \\
  Q2    & $1.26\pm0.20$& $2.7\pm0.3$   & 35.3/31  \\
\hline
\hline
\end{tabular}
\caption{Best fit parameters from simultaneous X-ray and optical
spectral analysis assuming a smoothed broken power law model (s=1,
from Granot \& Sari 2002) with the break energy between the optical
and the NIR bands (see \S2). We assumed different extinction curves to
evaluate rest frame host galaxy extinction $A_V$.
\label{tab:3}
}
\end{table}

\begin{acknowledgements} The authors thank the referee
F. J. Castander for his useful comments on the manuscript. 
      G.S. is supported by the 
      Research Training Network \emph {Gamma-Ray Bursts: an Enigma and a Tool}
	funded by the EU.
\end{acknowledgements}

\end{document}